\documentclass[preprint,amsmath,amssymb,nofootinbib]{revtex4-2}
\usepackage[utf8]{inputenc}
\usepackage{rotating}
\usepackage{hyperref}
\usepackage{array}
\usepackage{subfig}
\usepackage{float}
\usepackage{fancyhdr}
\usepackage{slashed}
\usepackage[inline]{enumitem}
\usepackage{dcolumn}
\usepackage{booktabs}
\usepackage{makecell}
\usepackage{placeins}
\usepackage[export]{adjustbox}
\usepackage{bm}
\usepackage{xcolor}
\usepackage{graphicx}
\usepackage{diagbox}
\usepackage{multirow}
\usepackage{ctable}
\usepackage[normalem]{ulem}
\usepackage[T1]{fontenc}
\definecolor{orange}{rgb}{1,0.5,0}

\begin{document}

\title{Direct constraints on the agnostic $\nu$SMEFT from heavy-neutrino searches at the LHC}

\author{Luc\'{\i}a Duarte}
\email{lucia.duarte@fcien.edu.uy}
\affiliation{Instituto de F\'{\i}sica, Facultad de Ciencias,
 Universidad de la Rep\'ublica \\ Igu\'a 4225, (11400) 
Montevideo, Uruguay.}

\author{Agustín Guillenea}
\email{aguillenea@fing.edu.uy}
\affiliation{Instituto de F\'{\i}sica, Facultad Ingenier\'ia,
 Universidad de la Rep\'ublica \\ Julio Herrera y Reissig 565, (11300) 
Montevideo, Uruguay,}
\affiliation{Instituto de F\'{\i}sica, Facultad de Ciencias,
 Universidad de la Rep\'ublica \\ Igu\'a 4225,(11400) 
Montevideo, Uruguay.}

\begin{abstract}

The observation of neutrino oscillations and masses motivates extensions of the Standard Model containing right-handed neutrinos. If additional new physics exists at scales beyond the direct reach of present experiments, its effects can be systematically described within the neutrino Standard Model Effective Field Theory ($\nu$SMEFT). In this framework, heavy neutral leptons provide a promising target for collider searches. In this work, we reinterpret a CMS search for heavy Majorana neutrinos in the same-sign dimuon plus two jets final state to constrain the parameter space of the dimension-six $\nu$SMEFT. Using a detailed recast of the experimental analysis, we present results for both a pure-agnostic benchmark and a benchmark incorporating neutrinoless double-beta-decay bounds. For heavy-neutrino masses in the range $200~\mathrm{GeV} \le m_N \le 15~\mathrm{TeV}$, the observed upper limits on the effective coupling strength range from $5.8\times10^{-7}\,\mathrm{GeV}^{-2}$ to $2.3\times10^{-6}\,\mathrm{GeV}^{-2}$. These results constitute the first experimental exclusion limits on the agnostic $\nu$SMEFT parameter space obtained from a dedicated recast of an LHC heavy-neutrino search.

\end{abstract}

\maketitle

\section{Introduction}{\label{intro}}
 

The discovery of the Higgs boson completed the experimental confirmation of the Standard Model (SM) mechanism of mass generation. Nevertheless, the origin of the tiny neutrino masses required to explain neutrino oscillation data remains unknown. A minimal extension of the SM capable of generating neutrino masses consists of introducing right-handed neutrino fields ($N_R$). Since these states are singlets under the SM gauge group, they can possess Majorana mass terms without violating gauge invariance. As a consequence, lepton-number-violating interactions may arise, and the physical neutrino spectrum can contain heavy Majorana neutrino states $N$. Such states are a generic prediction of many mechanisms of neutrino mass generation, including Type-I, Inverse, and Linear Seesaw constructions \cite{Minkowski:1977sc, Mohapatra:1979ia, Yanagida:1980xy, GellMann:1980vs, Schechter:1980gr, Mohapatra:1986bd, Malinsky:2005bi}.

Besides the mechanism responsible for neutrino mass generation, additional new physics may exist at energy scales well above the electroweak scale. In the absence of direct access to these degrees of freedom, their low-energy effects can be described in a model-independent way through the Standard Model Effective Field Theory (SMEFT). If heavy neutral leptons (HNLs) are part of the accessible particle spectrum, however, they should not be integrated out together with the ultraviolet states. In such a situation, the interactions of the HNLs with SM particles may receive sizeable contributions from higher-dimensional operators, potentially exceeding those induced solely by the neutrino-mass mixing mechanism. The resulting effective description is obtained by extending the SM field content to include right-handed neutrinos as explicit dynamical degrees of freedom. This framework, commonly referred to as the $\nu$SMEFT\footnote{Also denoted in the literature as SMNEFT, $N_R$SMEFT or $\nu_R$SMEFT.}, provides a systematic parametrization of the low-energy effects of ultraviolet physics in the presence of heavy neutrinos \cite{delAguila:2008ir,Aparici:2009fh,Liao:2016qyd,Bhattacharya:2015vja,Li:2021tsq}. Representative ultraviolet completions include Left-Right symmetric models, $Z'$-mediated scenarios and leptoquark models \cite{Mohapatra:1979ia,Mohapatra:1980yp,Mohapatra:1980qe,Deppisch:2019kvs,Saha:2026wgv,Bhaskar:2023xkm}. The correspondence between effective operators and ultraviolet completions has been studied systematically in Ref.~\cite{Beltran:2023ymm}.

Since its original formulation as an effective field theory including right-handed neutrinos \cite{delAguila:2008ir}, the $\nu$SMEFT framework has developed into a powerful and systematic approach to study heavy neutral leptons beyond the minimal mixing paradigm. On the phenomenological side, the framework has been extensively applied to the study of heavy-neutrino decays \cite{Duarte:2015iba,Duarte:2016miz} and collider signatures at present and future experiments, including both prompt and displaced final states \cite{Duarte:2016caz,Caputo:2017pit, Biekotter:2020tbd,Cottin:2021lzz,Mitra:2022nri,Duarte:2023tdw,Duarte:2025zrg,Beltran:2025ilg,Bolton:2025tqw,Mitra:2026kxt}. Complementary probes from low-energy observables, meson decays and neutrinoless double beta decay have also been investigated \cite{Bischer:2019ttk,Dekens:2020ttz,Beltran:2023nli,Fernandez-Martinez:2023phj,deVries:2024mla}. 

Despite this broad phenomenological program, direct constraints on the $\nu$SMEFT parameter space derived from dedicated recasts of LHC searches for heavy neutrinos remain largely unexplored. In this work, we intend to fill this gap by recasting a CMS search for same-sign dimuon plus dijet events \cite{CMS:2022hvh} in terms of the $\nu$SMEFT operator coefficients. Our analysis is performed within an agnostic benchmark in which the effective interactions are treated on equal footing, allowing for a model-independent exploration of the parameter space.

The paper is organized as follows: in Sec. \ref{sec:agnosticB} we define the theoretical setup, explaining the agnostic $\nu$SMEFT benchmark, and review the effective contributions to the same-sign dimuon plus dijet signal in Sec.\ref{sec:SSdimuon_vsmeft}. Then in Sec. \ref{sec:LHC_Bounds} we review the CMS search for this signal in terms of HNLs, describe in detail our recast procedure and discuss the obtained exclusion limits. We summarize and present our perspectives in Sec. \ref{sec:summary}.

\section{$\nu$SMEFT setup and benchmark scenarios}\label{sec:agnosticB}

\subsection{Minimal $\nu$SMEFT}

We start by extending the SM Lagrangian with a single right-handed neutrino $N_R$ with a Majorana mass term ($\sim M_N$) in \eqref{eq:LagSeesaw}. While this minimal setup cannot reproduce the two observed light-neutrino mass splittings and the measured leptonic mixing pattern, it captures the essential phenomenology of a heavy neutral lepton \cite{Fernandez-Martinez:2023phj}. Such a framework effectively describes scenarios in which additional heavy neutrino states $N_{2,3,...}$ are decoupled from the accessible observables relevant to our analysis. The renormalizable $d=4$ Lagrangian extension then reads 
\begin{eqnarray}\label{eq:LagSeesaw}
\mathcal{L}_{\rm d=4}= \overline{N_R} i\slashed{\partial} N_R - \left( \frac{M_N}{2} \overline{N^{c}_R} N_R + \sum_{\ell} Y_\ell ~ \overline{L_\ell}\tilde{\phi} N_R + \text{ h.c.}\right).
\end{eqnarray}
Upon diagonalization, this Lagrangian leads to one massive heavy $N$ and the three known light neutrinos $\nu_i$ (with masses $m_{\nu_i}\sim 0.1$ eV), which are all of Majorana nature. The active flavor $\ell=e,\mu,\tau$ neutrino eigenstates $\nu_{\ell L}$ contain some part of the heavy $N$ due to the mixing $V_{\ell N} = Y_\ell ~v/\sqrt{2} M_N$: 
\begin{equation}\label{eq:mixing}
    \nu_{\ell L}=\sum_{i=1}^{3}U_{\ell i}\nu_i+V_{\ell N}N.
\end{equation}
We can thus consider the right-handed sterile neutrino $N_R$ to be the right-handed component of the (only) heavy $N$. 
Since our goal is to probe new-physics effects beyond those generated by the minimal seesaw mechanism, we neglect the heavy neutrino mixing with the active states $V_{\ell N}$, taking into account both the already stringent experimental constraints and the naive theoretical seesaw relation $V_{\ell N}\lesssim \sqrt{ \frac{m_{\nu}}{M_{N}}}$, which results in suppressed interactions with the SM electroweak currents for $N$ masses above the GeV scale. 

\renewcommand{\arraystretch}{1.2}
\begin{table}[tbp]
\begin{adjustbox}{width=\textwidth,center} 
\centering
\begin{tabular}{| c| c  l | l | c | c |}
\firsthline
\textbf{Type}      &     \textbf{Operator}       &            & \textbf{Interactions}   & \textbf{Coupling}   \\ \hline\hline
$N$  mass $d=5$ & $\mathcal{O}^{d=5}_{N\phi}$ ($\mathcal{O}^{d=5}_{\rm Higgs}$)  & $(\bar{N}N^{c})(\phi^{\dagger} \phi)$  &  $ ~h N N$ and Majorana mass term  & $\alpha^{d=5}_{N\phi}$   \\ \hline 
Dipole $d=5$ & $\mathcal{O}^{(5)}_{NB}$  & $(\bar{N}_a \sigma_{\mu \nu} N^{c}_b) B^{\mu \nu}$, $a \neq b$ & $~$ Dipoles $~d_{\gamma}, d_Z$ & $\alpha^{d=5}_{NB}$ \\  \hline \hline
$h$-dressed mixing   & $\mathcal{O}^{(i)}_{LN\phi}$ ($\mathcal{O}_{\rm LNH}^\beta$)  & $(\phi^{\dag}\phi)(\bar L_i N \tilde{\phi})$  & $~$ Yukawa$+$doublet ($V_{\ell N}$. and $m_{\nu}$) & $\alpha^{(i)}_{LN\phi}$  \\ \hline 
$~$ Bosonic & $\mathcal{O}_{NN\phi}$ ($\mathcal{O}_{\rm HN}$) & $ i(\phi^{\dag}\overleftrightarrow{D_{\mu}}\phi)(\bar N \gamma^{\mu} N)$                                & $~$ Neutral current ($NNZ$) & $\alpha_{NN\phi}$ \\ 
$~$ Currents   & $\mathcal{O}^{(i)}_{Nl\phi}$ ($\mathcal{O}_{\rm HN\ell}^{\beta}$) & $i(\phi^T \epsilon D_{\mu}\phi)(\bar N \gamma^{\mu} l_i) $  & $~$ Charged current ($N l W$) & $\alpha^{(i)}_{Nl\phi}$ \\  \hline
$~$ Dipoles & $\mathcal{O}^{(i)}_{NB}$ ($\mathcal{O}_{\rm NB}$)  & $(\bar L_i \sigma^{\mu\nu} N) \tilde \phi B_{\mu\nu}$  &  $~$ One-loop level generated & $\alpha^{(i)}_{NB}/(16\pi^2)$ \\ 
$ $  & $\mathcal{O}^{(i)}_{NW}$ ($\mathcal{O}_{\rm NW}^\beta$)    & $(\bar L_i \sigma^{\mu\nu} \sigma^I N) \tilde \phi W_{\mu\nu}^I$ & $~~~d_{\gamma} , d_Z, d_W$ & $\alpha^{(i)}_{NW}/(16\pi^2)$ \\ \hline 
$ $  & $\mathcal{O}^{(i)}_{QNN}$ ($\mathcal{O}_{\rm QN}$)  &   $(\bar{Q_i} \gamma^\mu Q_i) (\bar{N} \gamma_\mu N)$  & $~ $ 4-fermion & $\alpha^{(i)}_{QNN}$   \\ \cline{2-3}\cline{5-5} 
$~$ 4-fermion N & $\mathcal{O}^{(i)}_{LNN}$ ($\mathcal{O}_{\rm LN}^{\beta}$) & $(\bar{L_i} \gamma^\mu L_i) (\bar{N} \gamma_\mu N)$  & $~$  vector- mediated  & $\alpha^{(i)}_{LNN}$   \\ \cline{2-3}\cline{5-5} 
$ $ &  $\mathcal{O}^{(i)}_{fNN}$ ($\mathcal{O}_{\rm ff}$) &   $(\bar f_i \gamma^{\mu}f_i) (\bar N \gamma_{\mu}N)$  & $~$ $f=u, d, l$  & $\alpha^{(i)}_{fNN}$   \\ \hline
 $~ $ 4-fermion CC & $\mathcal{O}^{(i, j)}_{duNl}$ ($\mathcal{O}_{\rm duN\ell}^{\beta}$) &  $ (\bar d _j \gamma^{\mu} u _j) (\bar N \gamma_{\mu} l_i)$ & $ ~$ 4-fermion vector- mediated  & $\alpha^{(i, j)}_{duNl}$    \\ \hline 
 $ $ & $\mathcal{O}^{(i, j)}_{QuNL} $ ($\mathcal{O}_{\rm QuNL}^\alpha$) & $(\bar Q _i u _i)(\bar N L_j)$ & $ ~$ 4-fermion & $ \alpha^{(i,j)}_{QuNL}$  \\ \cline{2-3}\cline{5-5} 
$ ~$ 4-fermion &  $\mathcal{O}^{(i, j)}_{LNQd} $   ($\mathcal{O}_{\rm LNQd}^\alpha$)  & $(\bar L_i N) \epsilon (\bar Q _j d _j)$  & $ ~$ scalar-mediated  & $\alpha^{(i,j)}_{LNQd}$ \\ \cline{2-3}\cline{5-5}
$~$  CC/NC & $\mathcal{O}^{(i, j)}_{QNLd}$ & $(\bar Q _i N)\epsilon (\bar L_j d_j)$  &  $ $ & $\alpha^{(i,j)}_{QNLd}$ \\ \cline{2-3}\cline{5-5} 
$ $ & $\mathcal{O}^{(i, j)}_{LNLl}$  ($\mathcal{O}_{\rm LNL\ell}^{\delta\beta}$) & $(\bar L_i N)\epsilon (\bar L_j l_j)$  & $ $ & $\alpha^{(i, j)}_{LNLl}$ \\ \lasthline
\end{tabular}
\end{adjustbox}
\caption{{Basis of $d=5$ and $d=6$ operators with a right-handed neutrino $N$ \cite{delAguila:2008ir, Liao:2016qyd}. Here $l_i$, $u _i$, $d _i$ and $L_i$, $Q _i$ denote the right-handed singlets and the left-handed $SU(2)$ doublets, respectively. The field $\phi$ is the scalar doublet, $B_{\mu\nu}$ and $W_{\mu\nu}^I$ are the $U(1)_{Y}$ and $SU(2)_{L}$ field strengths. Also $\sigma^{\mu \nu}=\frac{i}{2}[\gamma^{\mu}, \gamma^{\nu}]$ and $\epsilon=i\sigma^{2}$ is the anti-symmetric symbol in two dimensions. We follow the notation in \cite{delAguila:2008ir} and quote the names in \cite{Fernandez-Martinez:2023phj}.}\label{tab:Operators}}
\end{table}

This assumption allows us to focus on the impact of new physics that could arise from new heavy mediators at higher energy scales $\Lambda$. This new UV physics is captured in the $\nu$SMEFT set of effective operators $\mathcal{O}_\mathcal{J}$, constructed with SM fields and the right-handed neutrino $N_R$ and  consistent with the SM gauge symmetry $SU(2)_L \otimes U(1)_Y$\cite{delAguila:2008ir, Liao:2016qyd, Wudka:1999ax}, which lead to the following Lagrangian:
\begin{eqnarray}\label{eq:Lagrangian}
\mathcal{L}=\mathcal{L}_{SM}+\sum_{d=5}^{\infty}\sum_{\mathcal{J}} \frac{\alpha_{\mathcal{J}}}{\Lambda^{d-4}} \mathcal{O}_{\mathcal{J}}^{d}.
\end{eqnarray}
Here $d$ is the mass dimension of the operator $\mathcal{O}_{\mathcal{J}}^{d}$, $\alpha_{\mathcal{J}}$ are the effective (Wilson) couplings and the sum in $\mathcal{J}$ goes over all independent interactions at a given dimension $d$. We  list the $\nu$SMEFT operators up to $d=6$ in Tab. \ref{tab:Operators}. 

We have implemented the $\nu$SMEFT Lagrangian in the \texttt{FeynRules 2.3} package (see \cite{Zapata:2022qwo} for details) and given a discussion of the phenomenological role of each operator in \cite{Zapata:2023wsz}.

\subsection{Agnostic benchmark}\label{sec:agnostic}

Most studies of $\nu$SMEFT phenomenology are based on studying the impact of specific operators \cite{Butterworth:2019iff, Barducci:2022hll, Mitra:2024ebr, Beltran:2025ilg, Alonso:2025gzl} or take them to act separately at a time \cite{Fernandez-Martinez:2023phj, Beltran:2021hpq}. However, many of the operators in Tab. \ref{tab:Operators} can contribute directly to the heavy $N$ production at the LHC, and also modify the value of its total decay width \cite{Duarte:2016miz}. Thus, in a complementary strategy, we consider what we call an agnostic scenario, where we take into account the simultaneous effect of every dimension six operator in Tab. \ref{tab:Operators}\footnote{In this case we do not consider the $d=5$ operators, as they do not contribute to the processes studied at the LHC, nor to the decay of the heavy $N$, when only one right-handed neutrino state is present and the heavy-active mixings $V_{\ell N}$ are neglected.}. 

Our procedure consists of assigning a common numerical value to all effective coefficients, namely $\alpha_{\mathcal J}/\Lambda^2 \equiv \alpha/\Lambda^2$, for every operator $\mathcal{O}_{\mathcal J}$ appearing in Eq.~\eqref{eq:Lagrangian}. Under this assumption, the parameter space is reduced to two independent quantities: the heavy neutral lepton mass ($m_N$) and the overall effective coefficient $\alpha/\Lambda^2$, which controls the strength of the effective interactions and, consequently, the predicted observables.

We adopt this agnostic framework because generic ultraviolet completions are expected to generate several effective operators simultaneously rather than a single interaction term. Furthermore, matching at the new-physics scale and the subsequent renormalization-group evolution generally induce additional operators through operator mixing \cite{Chala:2020vqp, Beltran:2023ymm, Fernandez-Martinez:2023phj, Ardu:2024tzb}. As a result, scenarios in which only one effective operator is present are often not representative of realistic ultraviolet models. The assumption of a common effective coefficient therefore provides a simple and model-independent benchmark that captures the collective impact of the effective interactions while avoiding biases toward any particular ultraviolet realization. It should not be interpreted as a prediction of any specific ultraviolet completion, but rather as a convenient parametrization that allows us to characterize the sensitivity of the search in a reduced parameter space.

\subsection{$0\nu\beta\beta$-constrained benchmark}\label{sec:doblebetaBound}

Several studies have derived bounds on the $\nu$SMEFT Wilson coefficients $\alpha_{\mathcal J}$, or equivalently on the new-physics scale $\Lambda$, from direct and indirect searches for physics beyond the SM. However, most of these constraints are relevant for heavy neutral lepton masses below the range considered in this work ($m_N \lesssim m_W$) and are generally obtained under assumptions that differ from the agnostic benchmark adopted here \cite{Fernandez-Martinez:2023phj}. For HNL masses well below the electroweak scale, the appropriate effective description is provided by the $\nu$LEFT, whose operator basis is invariant under QCD and QED and includes sterile-neutrino interactions \cite{Chala:2020vqp,Datta:2020ocb,Ardu:2024tzb}.

Among the existing experimental constraints on the Majorana nature of heavy neutrinos, the most stringent ones typically arise from the non-observation of neutrinoless double beta decay ($0\nu\beta\beta$). These limits affect operators involving first-generation fermions. In the benchmark considered here, we impose the corresponding $0\nu\beta\beta$-bound on all effective operators contributing to the $udNe$ vertex, which are those associated with the coefficients $\alpha^{(1)}_{N l \phi}\; $ and $\alpha^{(1)}_{NW} $, contributing through the interchange of a $W$ boson, and the four-fermion $\alpha^{(1, 1)}_{duNl}, ~\alpha^{(1, 1)}_{QuNL}, ~\alpha^{(1, 1)}_{LNQd}, ~\alpha^{(1, 1)}_{QNLd}$.

More specifically, we require their coefficients to satisfy
\begin{equation}\label{eq:onubblimit}
\frac{\alpha_{0\nu\beta\beta}(m_N)}{\Lambda^2} = 3.2\times 10^{-8}\left(\frac{m_N}{100 ~ ~\rm{GeV}} \right)^{1/2},
\end{equation}
which we use as an upper limit for the relevant operator coefficients. Details of the derivation can be found in Refs.~\cite{Duarte:2016miz,Duarte:2014zea}. 

Thus, as a second benchmark, we consider a $0\nu\beta\beta$-constrained scenario. In contrast to the pure-agnostic benchmark introduced above, only the operators contributing to the $udNe$ vertex are assigned coefficients fixed by the  $0\nu\beta\beta$ limits in Eq.\eqref{eq:onubblimit}. All remaining effective coefficients are treated according to the agnostic prescription, i.e. they are assigned a common value $\alpha/\Lambda^2$. This benchmark allows us to assess the impact of the strongest low-energy constraints on the collider phenomenology of heavy neutral leptons \footnote{For more detailed studies of these limits see, for instance, Refs. \cite{Dekens:2020ttz,Beltran:2023ymm}}.

\section{Same-sign dimuons signatures in the $\nu$SMEFT}\label{sec:SSdimuon_vsmeft}

In this work we focus on the same-sign di-muons plus two jets signal, which has been thoroughly studied as the golden channel for discovery of a lepton number violating signal at the LHC \cite{CMS:2022hvh,ATLAS:2024rzi,CMS:2018jxx,CMS:2015qur, ATLAS:2015gtp,Atre:2009rg,delAguila:2007qnc}. The same-sign dilepton signal was the first one to be proposed to search for the dimension-six $\nu$SMEFT effective interactions at the LHC and the Tevatron \cite{delAguila:2008ir}. The authors originally considered the production of an on-shell $N$ in the hard-process $u \bar{d} \to N \ell^{+}$ followed by the decay $N \to \ell^{+} j j$. The $N$ production (and subsequent decay) can be realized by the $\mathcal{O}_{N l \phi}$ interaction, but also by the four-fermion operators $\mathcal{O}_{duNl}, \mathcal{O}_{QuNL}, \mathcal{O}_{LNQd}$ and $\mathcal{O}_{QNLd}$. This channel was also investigated in \cite{Duarte:2016caz}, focusing on a lighter heavy-$N$ mass regime $m_N< m_W$, where it produces a displaced signal which can be probed with muons forward-backward asymmetries. 
In \cite{Cirigliano:2021peb}, the authors constrain the contribution of the $\mathcal{O}_{N l \phi}$ operator to the Drell-Yan charged-current (CC)  $p\,p \to W^{\pm} \to \mu^{\pm} N \to \mu^{\pm} \mu^{\pm}\,j\,j$ process with the results of the previous CMS search for the same-sign dileptons signal \cite{CMS:2018jxx}, considering the HNL decays as $N \to \mu^{\pm} W^{\mp}$ by the seesaw mixing.  

\begin{figure}[tbh]
\centering
    \includegraphics[width=0.32\textwidth]{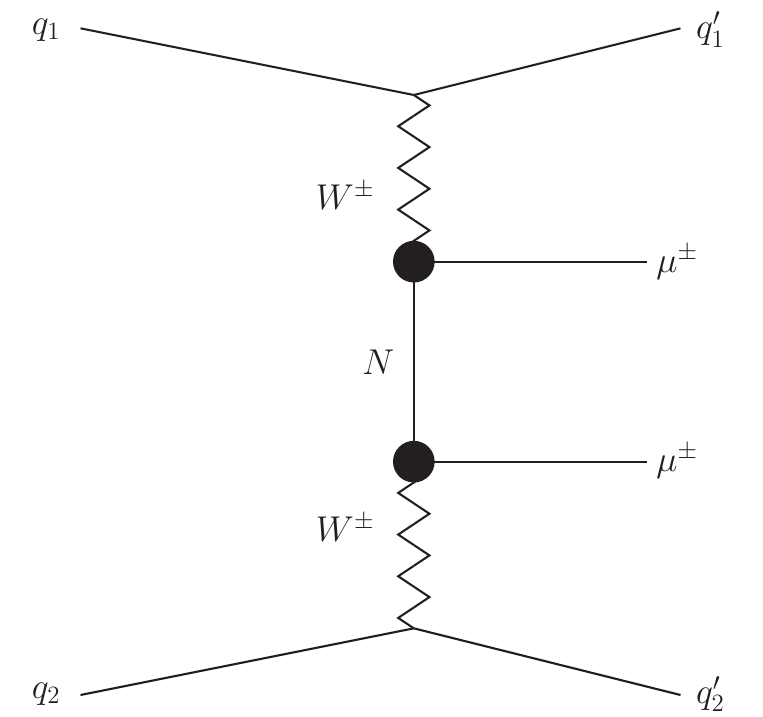}
    \includegraphics[width=0.32\textwidth]{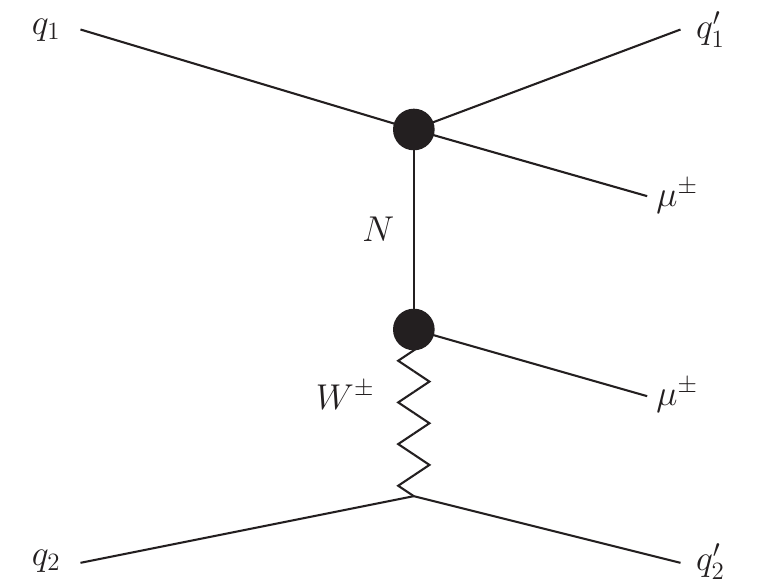}
    \includegraphics[width=0.32\textwidth]{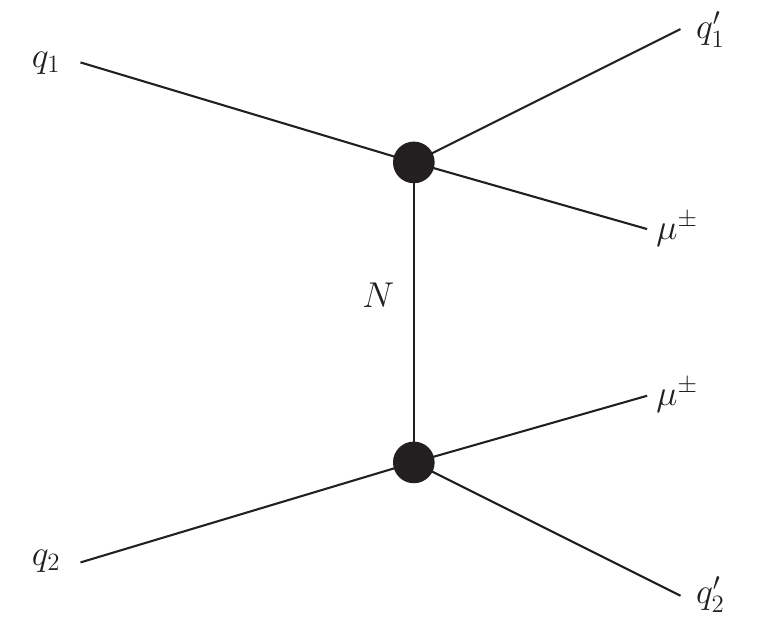}
\caption{Feynman diagrams for some realizations of the $p\,p\rightarrow \mu^{\pm}\,\mu^{\pm}\,j\,j$ process in the $\nu$SMEFT. Left: VBF production mediated by the $\mathcal{O}_{N l \phi}$ operator. Center: mixed production with $\mathcal{O}_{4f}$ (top vertex) and $\mathcal{O}_{N l \phi}$ (bottom vertex). Right: production with two (possibly different) $\mathcal{O}_{4f}$ interactions.}
\label{fig:sig_diag_neff}
\end{figure}

There are many possible contributions from the effective operators in Tab. \ref{tab:Operators} to the $p\,p\rightarrow \mu^{\pm}\,\mu^{\pm}\,j\,j$ signature. In Fig. \ref{fig:sig_diag_neff} we show some of the tree-level topologies for different realizations of the same-sign di-muons production. 
Although the CMS search recasted here \cite{CMS:2022hvh} was originally targeted for the $W^{\pm}W^{\pm} \to \mu^{\pm} \mu^{\pm}$ boson fusion sub-process from the Type-I seesaw model \cite{Fuks:2020att}, we find many contributions from effective interactions that do not mimic this VBF signature. The diagram on the left in Fig. \ref{fig:sig_diag_neff} is the only one that really resembles the searched signal. There, the heavy $N$ produced from $W$ bosons in a t-channel configuration mediated by the right-handed charged-current operator $\mathcal{O}_{N l \phi}$ allows for two-unit lepton number violation\footnote{The same diagram can be drawn with the one-loop-level generated interaction from the $\mathcal{O}_{NW}$ operator in each vertex. We do not consider it in our numerical simulation, because it is suppressed by a squared loop factor $(16\pi^2)^{-2}$ in comparison with the $\mathcal{O}_{N l \phi}$ vertex.}. 

Here we also consider tree-level diagrams where the $N$ is not produced on-shell, with contributions form the right-handed CC $\mathcal{O}_{N l \phi}$ and the four-fermion operators $\mathcal{O}_{duNl}, \mathcal{O}_{QuNL}$, $ \mathcal{O}_{LNQd}$ and $\mathcal{O}_{QNLd}$, as in the center and right diagrams in Fig. \ref{fig:sig_diag_neff}. In the agnostic scenario, all dimension-six $\nu$SMEFT operators contribute simultaneously to the signal, and interference effects among the different amplitudes are fully included. Since these operators generate a variety of production topologies that are not necessarily characterized by the VBF-like kinematics targeted by the CMS analysis, a substantial fraction of the $\nu$SMEFT signal is removed by the selection cuts defining the signal region.

In the context of the Type-I seesaw based search originally proposed in \cite{Fuks:2020att}, the interference between the searched VBF process $p \, p \to W^{\pm}W^{\pm} \,j\,j  \to \mu^{\pm} \, \mu^{\pm} \,j\,j $ with a  t-channel $N$ and the s-channel process $q \bar{q'} \to W^{(*)\pm} \to \mu^{\pm} N  \to \mu^{\pm} \, \mu^{\pm} \,j\,j $ was neglected. This is justified in that context, because for $m_N <$ TeV scale, the $N  \to W^{(*)\mp} \mu^{\pm} \to \mu^{\pm} \,j\,j $ splitting chain is suppressed by the narrow width approximation. The search imposes a condition on the invariant mass of the jets pair $m_{j j}$ to be sufficiently high to suppress the contibutions from this decay chain. However, in the agnostic $\nu$SMEFT scenario, this is no longer the case, because the $N$ decay to a lepton and two jets is not  dominated by the $W \ell$ channel only, but also receives important contributions from the four-fermion interactions, as was originally pointed out in \cite{Duarte:2016miz}.   

\section{Recast of the CMS search}\label{sec:LHC_Bounds}

\subsection{CMS search strategy}\label{sec:CMS-Search}
 
We recast the CMS search for heavy Majorana neutrinos reported in Ref.~\cite{CMS:2022hvh}, entitled \textit{Probing Heavy Majorana Neutrinos and the Weinberg Operator through Vector Boson Fusion Processes in Proton-Proton Collisions at $\sqrt{s}=13$ TeV}. This analysis was the first LHC search to exploit vector boson fusion processes to probe heavy Majorana neutrinos and the Weinberg operator. It was performed in the same-sign dimuon and dijets final state using a proton-proton collision data set recorded at $\sqrt{s}=$13 TeV, collected with the CMS detector, corresponding to a total integrated luminosity of 138 fb$^{-1}$. No significant deviation from the prediction of the SM was observed and the results were translated into upper limits on the mixing parameter $|V_{\mu N}|^2$ for heavy-neutrino masses between 50 GeV and 25 TeV.

The search was originally proposed in \cite{Fuks:2020att} and targets events with two forward jets from vector boson fusion (VBF), two same-sign muons, and little missing transverse momentum, the latter being a smoking-gun signature of lepton number violation.
The heavy Majorana neutrinos signal is modeled in the phenomenological Type-I Seesaw \cite{Atre:2009rg, delAguila:2008cj} implemented for numerical simulation in \cite{Degrande:2016aje, Alva:2014gxa} in the \texttt{FeynRules} package \cite{Alloul:2013bka, Degrande:2011ua}, and then interphased with the Monte Carlo (MC) event generator \texttt{MadGraph5\_aMC@NLO}. The signal process is $p\,p\rightarrow \mu^{\pm}\,\mu^{\pm}\,j\,j$. In this context, the two same-sign muons can be produced in $W^{\pm}W^{\pm}\to \mu^{\pm}\,\mu^{\pm}$ scattering \cite{Dicus:1991fk}, mediated by a heavy $N$ in the t-channel .  

Because the cross section of t-channel processes is less sensitive to the mass of the intermediate particle compared to s-channel quark-antiquark annihilation, the VBF process can complement searches for heavy Majorana neutrinos at the TeV mass scale, as its cross section decreases more slowly with increasing $N$ mass compared to the values from s-channel production. Here the $N$ is non-resonant, and it cannot behave as a long-lived particle, despite its decay width value \cite{Fuks:2020att}, which prevents the appereance of displaced signatures.

The search analyzes final states with two well identified same-sign muons, with pseudorapidities $|\eta_{\mu}|< 2.4$ and $p_{T \mu} > 30$ GeV, and containing at least two jets with $|\eta_{j}|< 4.7$ and $p_{T j}>30$ GeV. The two jets with the highest $p_T$ are chosen as the VBF candidates, so they are required to have a large invariant mass $m_{j_1 j_2}> 750$ GeV and pseudorapidity separation $|\Delta \eta_{j_1 j_2}|> 2.5$. Also, by momentum conservation, the muons are expected to be produced near the transverse plane. Thus, a rapidity gap is expected between the muons and the VBF jets. This gap is quantified using the Zeppenfeld variable of a lepton $\ell$, $\mathcal{Z}_\ell = |\eta_\ell -(\eta_{j_1}+\eta_{j_2})/2| / |\Delta \eta_{j_1 j_2}|$ \cite{Rainwater:1996ud}. A cut on max$(\mathcal{Z}_\ell)< 0.75$ is imposed for the muons, as it measures how much they are separated from the average pseudorapidity value of the VBF jets. 

In addition, events with loosely identified muons with $|\eta_\mu|<2.5$ ($\mu$-veto), or electrons with $|\eta_e|<2.4$ and $p_T> 10$ GeV ($e$-veto) are rejected, together with events with hadronically decaying taus with $|\eta_\tau|< 2.3$ and $p_{T \tau}>20$ GeV  (had-$\tau$ veto). These vetos are intended to suppress contributions from $W^\pm Z$ production. Finally, events where at least one $b$-tagged jet with $|\eta|< 2.5$ and $p_T>20$ GeV is found are rejected  (b-jet veto), to avoid nonprompt-lepton backgrounds.
As signal muons are expected to be produced more back-to-back than in background processes, the signal region in the CMS analysis is separated into two bins with $\Delta\phi_{\ell\ell} \lessgtr 2$. 

The study employs the quotient between the scalar sum of the transverse momenta of every jet with $p_{T j }>30$ GeV and $\left| \eta_j \right|<4.7$, $H_T=\sum p_T^{(jets)}$ and the transverse momentum of the leading muon, $p_T^{\mu_1}$ as the discriminating variable. This observable ($H_T/p_T^{\mu_1}$) measures the hadronical activity compared to the leptonical activity in each event, and is expected to be less than 1 for the majority of the signal events, indicating that on an
event-by-event basis more transverse momentum is carried by the leading muon than in all jets combined, which is expected for a VBF production process \cite{Fuks:2020att, CMS:2022hvh}. 

The statistical analysis, based on the number of measured data $n$, expected background $n_b$ and signal $n_s$ events in the signal region in the $H_T/p_T^{\mu_1}$ bins for both $\Delta\phi_{\ell\ell} \lessgtr 2$ regions yields the upper limits at the 95\% C.L. in the mixings $|V_{\mu N}|^2$. All data required for the statistical analysis is made public in \cite{hepdata.130825}. 

\subsection{Event simulation and recast procedure}\label{sec:Our_recast}

In order to recast the CMS search results and interpret them in terms of the $\nu$SMEFT effective interactions, we simulate signal events and process them to replicate the CMS analysis, obtaining our expected number of signal events in the signal region, and perform the statistical analysis that allows us to set limits on the agnostic $\nu$SMEFT parameter space. We will now describe this procedure in detail. 

\begin{figure}[t]
\centering
    \includegraphics[width=0.94\textwidth]{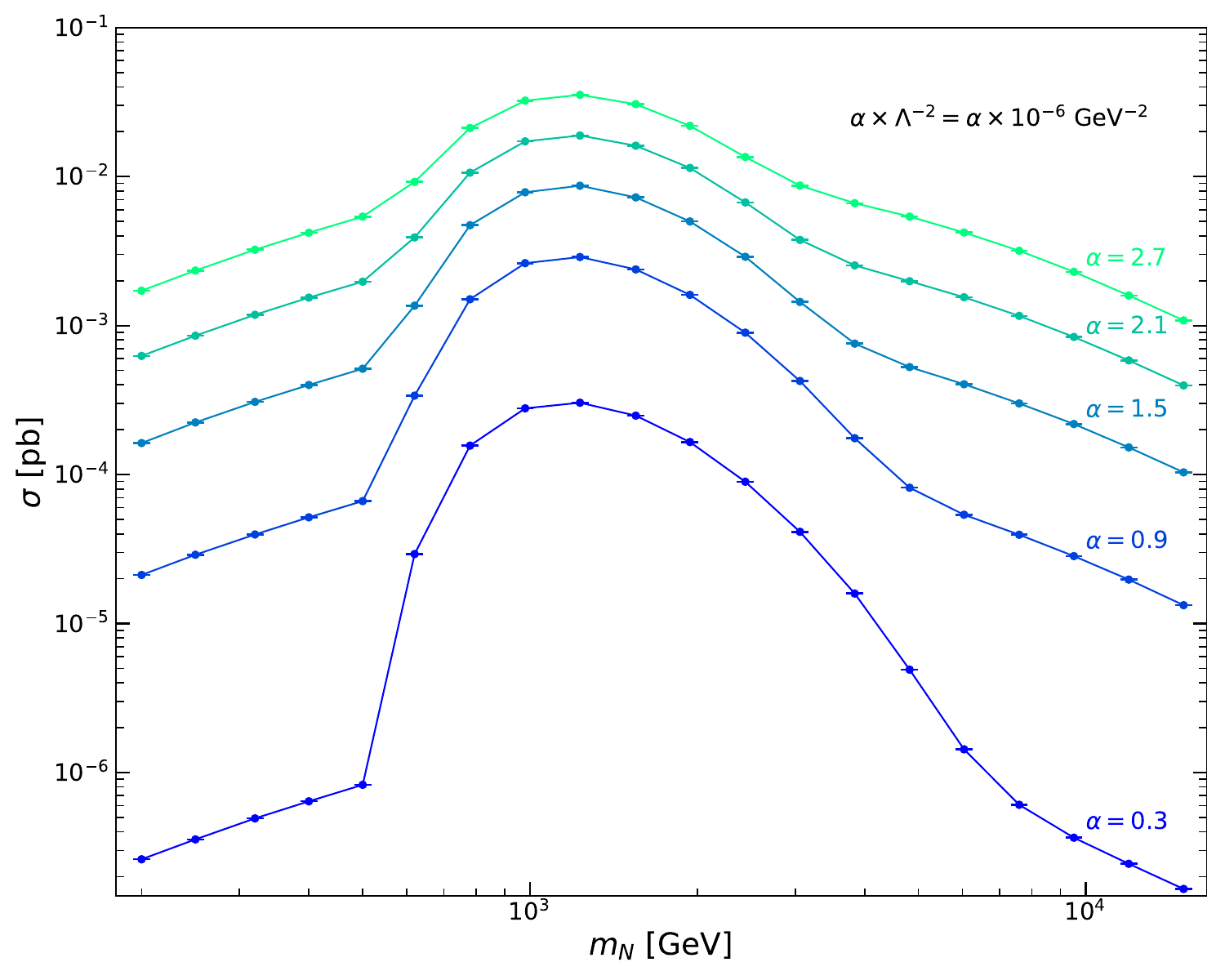}
\caption{Parton-level cross sections for the process $p \, p \to \mu^{\pm} \, \mu^{\pm} j j $ in the agnostic $\nu$SMEFT scenario at $\sqrt{s}=13$ TeV. The curves correspond to different values of the common effective coupling $\alpha/\Lambda^2$. The cross sections are computed using tree-level matrix elements and include the generation-level kinematic cuts described in the text.}
\label{fig:xsecVSmN}
\end{figure}

We generate signal events with the numerical implementation of the $\nu$SMEFT interactions in the agnostical scenario, using the UFO output for our FeynRules implementation, the Neff6 model \cite{Zapata:2022qwo}. The signal process $p\,p\rightarrow \mu^\pm \,\mu^\pm \,j\,j$ is generated with \texttt{MadGraph5\_aMC@NLO 3.6.2} \cite{Alwall:2014hca} considering tree-level Feynman diagrams\footnote{We do not include the one-loop generated interaction $\mathcal{O}_{NW}$ in the signal events generation.}, and colliding energy $\sqrt{s}=13$ TeV, with the \texttt{NNPDF31\_lo\_as\_0130} parton distribution functions from the  NNPDF collaboration \cite{NNPDF:2017mvq}.

We scan the agnostic $\nu$SMEFT parameter space by generating signal events on a grid of 20 heavy-neutrino mass values logarithmically spaced in the range $0.2 \,\text{TeV}< m_N < 15 \, \text{TeV}$ and 13 values of the effective coupling $\alpha/\Lambda^2$ linearly spaced between $10^{-7}$ GeV$^{-2}$ and $10^{-6}$ GeV$^{-2}$. 
To improve the efficiency of event generation, we impose parton-level cuts compatible with the subsequent analysis selection requirements: $\left|\eta_{\ell} \right|< 2.4 \, \text{for} \, \ell=e, \mu$,  $\left|\eta_{j} \right|<2.7$ and  $p_{T j,\ell }>30$ GeV, $\Delta R_{\ell\ell} = \Delta R_{jj} = \Delta R_{j\ell} = 0.4$ together with invariant-mass requirements $m_{\ell\ell}>20$ GeV and $m_{jj}>500$ GeV, and a minimum jets pseudorapidity separation $\left|\Delta\eta_{jj} \right|>2.5$.
These generation-level cuts significantly improve the Monte Carlo efficiency while preserving compatibility with the final signal-region selection. Representative parton-level cross sections obtained throughout the parameter-space scan are shown in Fig.~\ref{fig:xsecVSmN}.

Parton-level events are interfaced with \texttt{Pythia 8.313} \cite{10.21468/SciPostPhysCodeb.8, 10.21468/SciPostPhysCodeb.8-r8.3, Sjostrand:2014zea} for parton showering and hadronization, and we perform a fast detector simulation using \texttt{Delphes 3.5.0} \cite{deFavereau:2013fsa, Selvaggi:2014mya, Mertens:2015kba}, with the embedded CMS card. These events are analyzed with \texttt{MadAnalysis5 1.10.16} \cite{Araz:2023axv, 10.21468/SciPostPhys.14.1.009, Araz:2021akd, Araz:2020lnp, Araz:2019otb, Conte:2018vmg, Dumont:2014tja, Conte:2014zja, Conte:2012fm}.

\begin{figure}[t]
\includegraphics[width=0.9\textwidth]{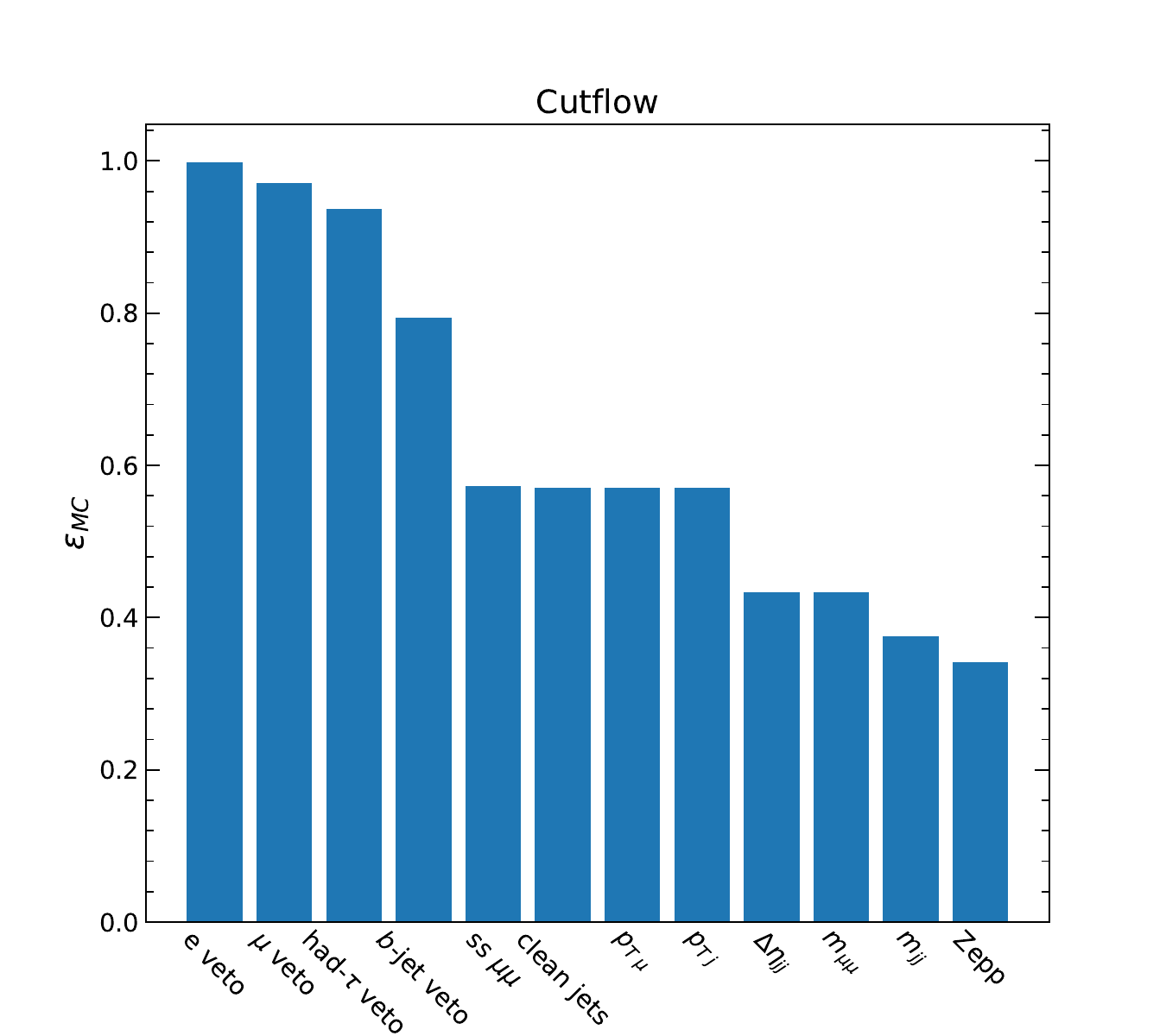}
\centering
\caption{Cutflow of the CMS signal-region selection applied to the $\nu$SMEFT signal samples. The cumulative efficiency $\varepsilon_{MC}$ is shown after each analysis requirement, including the lepton and b-jet vetoes, the same-sign dimuon and jet selections, the VBF-inspired kinematic cuts, and the Zeppenfeld variable requirement. Efficiencies are computed with respect to the total number of generated Monte Carlo events after generation-level preselection cuts and are averaged over all simulated signal points.}
\label{fig:cutflow}
\end{figure}

On top of the above generation cuts (or pre-selection cuts) we perform the analysis described in the previous section \ref{sec:CMS-Search}. The flowcut chart, showing the selection efficiencies for signal events to be in the signal region defined by the CMS analysis, is shown in Fig. \ref{fig:cutflow}. These are the number of selected MC events over the total MC generated events, which already include the pre-selection or generation cuts ($\varepsilon_{MC}$). The numbers displayed are the averaged values across all signal points generated in the parameter space grid. 

The electron, muon, hadronic-tau and b-tagged jets vetoes are followed by the selection of events with two well-identified same-sign muons (ss-$\mu \mu$) with pseudorapidities $|\eta_{\mu}|< 2.4$ and $p_T>10$ GeV and at least two well-identified jets (clean jets) with $|\eta_{j}|< 4.7$ and $p_T>15$ GeV. Next, events with both muons with $p_{T \, \mu}> 30$ GeV are selected ($p_{T \, \mu}$) and the two leading jets are required to have $p_{T \, j }> 30$ GeV  ($p_{T \, j}$). These jets are required to have a minimum separation $|\Delta \eta_{jj}|> 2.5$  ($\Delta \eta_{j j}$). The invariant mass of the muons pair is required to be $m_{\mu \mu}> 20$ GeV ($m_{\mu \mu}$) and jets with $m_{j j }> 750$ GeV ($m_{j j}$) are kept. The last cut is on the maximum value of the Zeppenfeld variable $max(\mathcal{Z}_\ell)< 0.75$ (Zepp). 

\subsection{Statistical interpretation}\label{sec:statistical}

We now turn to the statistical analysis used to constrain the model. We perform our statistical test taking into account all the events that pass the signal-region cuts, irrespective of the values of the discriminating variables. We do so to simplify the analysis, given that in the search the bining in both variables is done to better fit the background and enhance signal sensitivity. We unify two regions that are disjoint in the original search, distinguished by the angular separation in the transverse plane of the detector of the muon pair in the final state. In turn, in each of these regions, we unify the bins in the variable $H_T/p_T^{\mu_1}$, which correspond to values of 0-1,
1-2, 2-5 and > 5 for the high $\Delta\phi_{{\mu \mu}}$ region, and values of 0-2, 2-5 and > 5 for the low $\Delta\phi_{{\mu \mu}}$ region. Taking the observed number of events $n$ and expected background events $b$ in the signal region of \cite{CMS:2022hvh, hepdata.130825}, we consider a counting experiment with a Poisson likelihood function and calculate the
upper number of signal events $s^{up}$ consistent at $=95\%$ confidence level with the observation of $n$ events and a background prediction of $b$ events.
 Our implementation follows the PDG review on Statistics \cite{ParticleDataGroup:2024cfk} and Appendix B in \cite{Magill:2018jla}.
 We then consider, for every parameter-space point $(m_N, \alpha/\Lambda^2)$ in the simulated grid, the number of signal events that are predicted in the analysis signal region as 
 \begin{equation}
    s = \mathcal{L}\, \sigma_{MC}\, \varepsilon_{MC},
\label{eq:signalevents}
\end{equation}
where $\sigma_{MC}$ and $\varepsilon_{MC}$ are the MC-level cross-section obtained for each simulated point and the selection efficiencies, respectively. We considered an integrated luminosity of 138 fb$^{-1}$, equal to the total luminosity evaluated in the original search. This procedure allows us to exclude the parameter-space region in which the interpolated predicted number of signal events $s$ exceeds the $s^{up}$ value derived from the experiment. 

\subsection{Results}\label{sec:results}

\begin{figure}[t]
\includegraphics[width=0.9\textwidth]{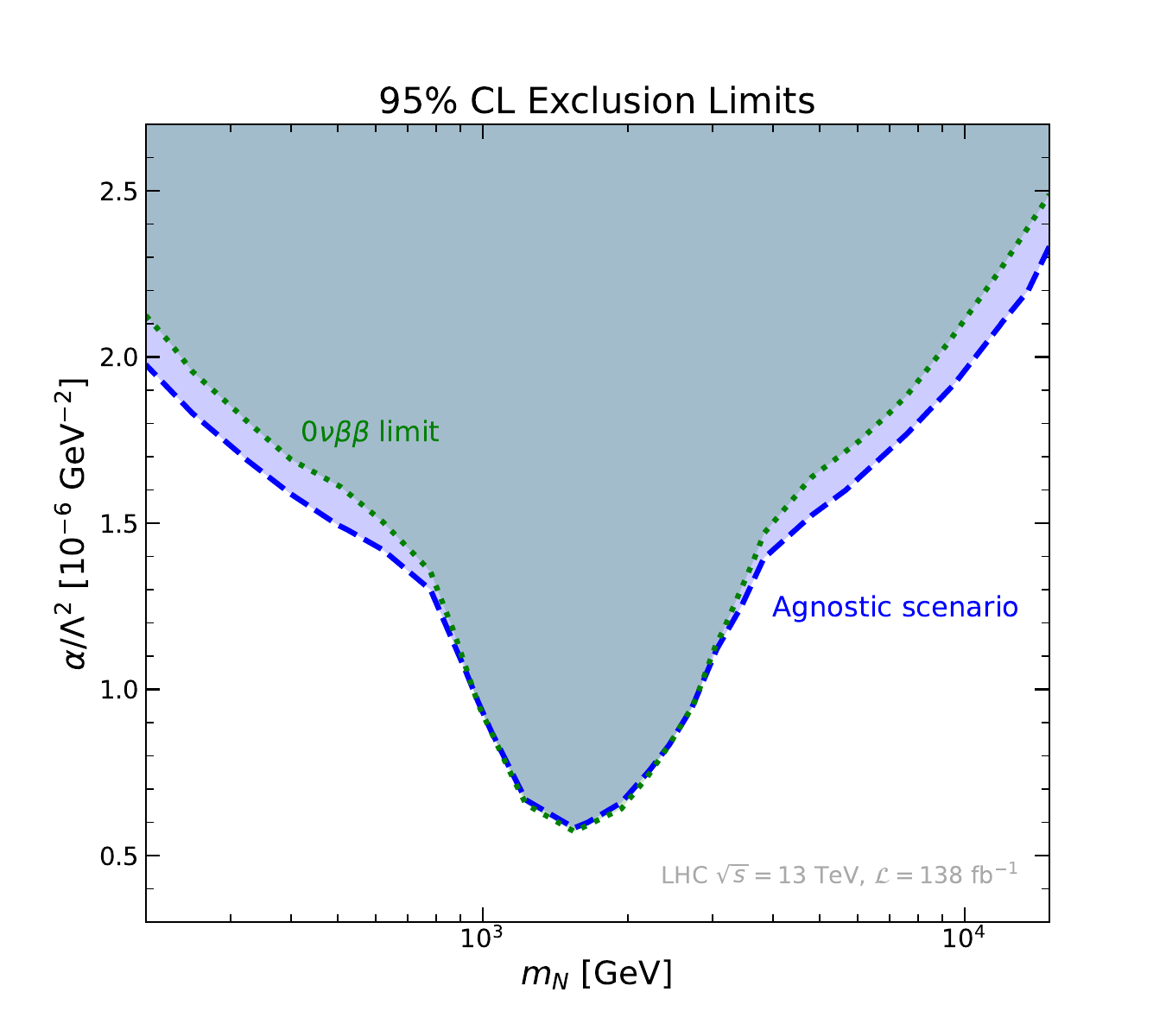}
\centering
\caption{Expected 95\% C.L. exclusion limits in the agnostic $\nu$SMEFT parameter space obtained from a recast of the CMS VBF search for heavy Majorana neutrinos \cite{CMS:2022hvh}. The excluded regions correspond to parameter points for which the predicted signal yield exceeds the upper limit derived from the CMS analysis. Results are shown for the fully agnostic benchmark (blue dashed contour) and for the $0\nu\beta\beta$-constrained benchmark (green dotted contour).}
\label{fig:Neff_contour}
\end{figure}

The resulting exclusion limits are shown in Fig. \ref{fig:Neff_contour}. The shaded areas correspond to the parameter-space regions where the interpolated expected number of signal events $s$ exceeds the upper allowed value $s^{up}$, and are thus excluded by the analysis. The blue dashed curve shows the limit for the simple agnostic benchmark scenario, where all the effective couplings $\alpha/\Lambda^2$ are set to the same numerical value for every operator and for every fermion family. We also show in dotted green the limit obtained in the case of the $0\nu\beta\beta$-constrained benchmark, where we set the values of the couplings of the operators contributing to neutrinoless double beta decay to be equal to the upper bound as explained in Sec. \ref{sec:doblebetaBound}.

The exclusion contours shown in Fig.~\ref{fig:Neff_contour} exhibit a characteristic U-shaped behaviour. This feature can be understood from the dependence of the signal yield on both the production cross section and the event-selection efficiency. Since the amplitudes generated by the dimension-six operators scale linearly with the effective coupling, the signal cross section behaves approximately as $\sigma \propto (\alpha/\Lambda^2)^2$.
The expected number of signal events after selection can therefore be written as
\begin{equation}
s=
\left(\frac{\alpha}{\Lambda^2}\right)^2
\tilde{\sigma}(m_N)\,
\varepsilon \,
\mathcal{L},
\end{equation}
where $\tilde{\sigma}(m_N)$ denotes the cross section evaluated for $\alpha/\Lambda^2=1$. For a fixed upper limit on the allowed signal yield, the excluded coupling scales approximately as $(\tilde{\sigma}(m_N)\,\varepsilon)^{-1/2}$. 
The exclusion contour therefore reflects the behaviour of the selected signal rate as a function of the HNL mass. The strongest limits are obtained around $m_N\sim1.5$ TeV, where the selected signal yield reaches its maximum. For both lower and higher masses, the number of selected signal events decreases, requiring larger values of the effective coupling to remain observable. As a consequence, the minimum excluded value of the effective coupling is approximately $\alpha/\Lambda^2 \simeq 5.8\times10^{-7}\,\mathrm{GeV}^{-2}$. This constitutes the first direct exclusion limit on the agnostic $\nu$SMEFT parameter space derived from a reinterpretation of LHC heavy-neutrino search data.

The U-shaped exclusion contour closely resembles the behaviour observed in the original CMS limits obtained within the Type-I seesaw model \cite{CMS:2022hvh}. This reflects the fact that the sensitivity of the search is maximized in an intermediate mass region, while it decreases for both lighter and heavier HNLs.
However, the position of the minimum is shifted with respect to the CMS interpretation, reflecting the richer signal composition present in the agnostic $\nu$SMEFT scenario. Although the CMS analysis was optimized for VBF production of Majorana neutrinos within the Type-I seesaw framework, the reinterpretation performed here demonstrates that the search also exhibits significant sensitivity to the broader class of LNV interactions described by the $\nu$SMEFT.

Remarkably, imposing the $0\nu\beta\beta$ constraints on the subset of operators contributing to neutrinoless double beta decay only produces a modest modification of the exclusion contour. The modest impact of the $0\nu\beta\beta$ constraint can be understood from the flavour structure of the benchmark. While the neutrinoless double beta decay bound affects the subset of operators involving first-generation fermions that contribute to the $udNe$ vertex, the same-sign dimuon signal also receives contributions from operators involving second-generation leptons, whose coefficients remain unconstrained. As a consequence, the signal rate is only partially reduced, leading to exclusion contours that remain close to those obtained in the fully agnostic scenario. This behaviour suggests that the sensitivity of the search is not exclusively driven by the subset of operators constrained by neutrinoless double beta decay, but rather receives contributions from several effective interactions entering the $\nu$SMEFT description. The proximity of the two exclusion contours demonstrates that current LHC searches retain substantial sensitivity to the agnostic $\nu$SMEFT framework even after imposing the strongest available low-energy constraints.

The fact that the derived limits correspond to effective couplings of order $10^{-6}\ \mathrm{GeV}^{-2}$ suggests that the associated $\nu$SMEFT contributions may be phenomenologically relevant in the same region of parameter space traditionally explored through heavy-light neutrino mixing. A quantitative comparison, as well as the study of the interplay between effective interactions and mixing-induced processes, is left for future work.

\section{Summary and Perspectives}\label{sec:summary}

The Standard Model Effective Field Theory extended with sterile neutrinos ($\nu$SMEFT) provides a systematic framework for describing the low-energy effects of ultraviolet new physics in scenarios containing heavy neutral leptons.

The phenomenology of particular operators in this broad framework has been studied intensively in the last decade, covering reinterpretations of existing searches at low-energy experiments, meson decays and hadron-collider signatures involving mostly interactions with Higgs bosons. In addition, new searches have been proposed for the LHC and the future HL-LHC, focusing on displaced signals. Prospective sensitivity studies have also been performed for future $e^{+}e^{-}$ and $e^{-} \,p$ colliders. Most phenomenological analyses have focused on individual operators considered separately.

Here we explore an alternative strategy, considering what we call an agnostic $\nu$SMEFT benchmark, where we include the simultaneous effect of different dimension- six effective operators. This agnostic approach captures the simultaneous contribution of multiple effective interactions and naturally includes the interference among the corresponding amplitudes.

We perform a dedicated recast of a CMS search for HNLs in the same-sign dimuons and dijets final state, which covers the high-mass regime \cite{CMS:2022hvh}. Signal events are generated in the agnostic $\nu$SMEFT scenario, and the CMS cut-based analysis strategy is implemented. A statistical interpretation based on the observed number of events and the expected background in the signal region then allows us to derive bounds directly on the fully-agnostic $\nu$SMEFT parameter space $(m_N, \alpha/\Lambda^2)$.  

The most stringent limits are reached for $m_N \sim 1.5$ TeV, excluding effective couplings down to $\alpha/\Lambda^2 \simeq 5.8\times 10^{-7}\,\mathrm{GeV}^{-2}$. This constitutes the first direct collider exclusion of the agnostic $\nu$SMEFT parameter space derived from an LHC heavy-neutrino search.

We find that the CMS VBF analysis is sensitive not only to the t-channel heavy-neutrino exchange topology for which it was originally designed, but also to a broader class of $\nu$SMEFT topologies that populate the signal region after the selection cuts. The reinterpretation preserves the characteristic mass dependence of the original VBF search while providing, for the first time, direct constraints on the agnostic $\nu$SMEFT parameter space. A remarkable feature of the result is the close agreement between the fully agnostic benchmark and the $0\nu\beta\beta$- constrained scenario, which imposes the strongest available low-energy constraints on first-generation operators.  

A particularly interesting extension of the present analysis would be the simultaneous inclusion of mixing-induced and $\nu$SMEFT contributions. Since both mechanisms can generate the same-sign dimuon signature, their combined effects and possible interference may reveal novel phenomenological features absent in either framework taken in isolation. Another important direction for future work is the disentanglement of the contributions from individual $\nu$SMEFT operators and the study of their interference patterns. Such analyses could clarify which effective interactions are most strongly constrained by current collider searches and facilitate the connection between experimental observables and specific ultraviolet completions.

\acknowledgments

We thank PEDECIBA, ANII and CSIC (Uruguay) for financial support. A. Guillenea gratefully acknowledges Sitian Qian, Andrea Massironi, Giulia Lavizzari, and Jie Xiao from the CMS Collaboration for the training provided during the CERN Summer Student Programme 2024, which was instrumental for the development of this work.

\bibliographystyle{bibstyle}
\bibliography{Bib_N_6_26}

\end{document}